\documentclass[showpacs,aps,pra,twocolumn]{revtex4}
\usepackage{graphicx,epsfig}
\usepackage{times}
\usepackage{amsmath,amssymb}

\newlength\figurewidth
\newcommand{\fig}{Fig.}
\newcommand{\figs}{Figs.}

\newcommand{\rem}[1]{}

\newcommand{\neff}{n_{\mbox{\footnotesize eff}}}

\begin{document}

\title{Unidirectional light emission from high-$Q$ modes in optical
  microcavities 
}
  \author{Jan Wiersig}
  \affiliation{Institut f{\"u}r Theoretische Physik, Universit{\"a}t Bremen,
  Postfach 330 440, D-28334 Bremen, Germany}
  \author{Martina Hentschel}  
  \affiliation{Institut f{\"u}r Theoretische Physik, Universit{\"a}t
  Regensburg,  D-93040 Regensburg, Germany}
  \affiliation{ATR Wave Engineering Laboratories, 2-2-2
  Hikaridai, Seika-cho, Soraku-gun, Kyoto 619-0228, Japan}
\date{\today}

\begin{abstract}
We introduce a new scheme to design optical microcavities supporting 
high-$Q$ modes with unidirectional light emission.  This is 
achieved by coupling a low-$Q$ mode with unidirectional emission to a 
high-$Q$ mode. The coupling is due to
enhanced dynamical tunneling 
near an avoided resonance crossing. Numerical results for a microdisk with a
suitably positioned air hole demonstrate the feasibility and the 
potential of this concept.   
\end{abstract}
\maketitle

Confinement and manipulation of light in microcavities is important for a wide
range of research areas and applications, e.g., cavity quantum electrodynamics
or novel light sources~\cite{Vahala03}.  
A key quantity characterizing a  cavity mode
is its quality factor $Q = \omega/\Delta\omega$, where $\omega$ is the mode
frequency and $\Delta\omega$ is the linewidth. A large
$Q$-factor is a basic requirement for low threshold lasing,
high sensor sensitivity, narrow wavelength-selective filtering, and strong
light-matter interaction. Whispering-gallery modes (WGMs) in
microdisks~\cite{MLSGPL92}, microspheres~\cite{CLBRH93}, and
microtori~\cite{IGYM01} have {\it ultra-high $Q$-factors}. For  
state-of-the-art semiconductor microdisks, the record $Q$-factor is
$>3.5\times 10^5$~\cite{SBJBP05,BSBP04}.    
The applicability of those cavities as microlasers and single-photon sources
is, however, limited by isotropic light emission. The best directionality
so far is provided by VCSEL-micropillars; see, e.g., Ref.~\cite{RBGG01}. The 
emission is {\it unidirectional} at the cost of a reduced $Q$-factor,
typically well below $10^4$.  
With present technology, there is a trade-off between $Q$-factor and
directionality.    
  
This dilemma remains when breaking the rotational symmetry of a microdisk. 
Shape deformation~\cite{LSMGPL93,ND97} allows improved directionality of 
emission due to {\it refractive escape}, but the $Q$-factors are significantly
spoiled.  
Unidirectional emission has been reported for rounded
triangles~\cite{KLRK04} with $Q \approx 35$ and for spiral-shaped
disks~\cite{KTMJCC04}. In the latter case, a strong degradation of the cavity
$Q$ allows lasing operation only for spirals of the size of conventional edge
emitting lasers.  

Another approach is to break the symmetry by modifying the
evanescent leakage (the optical analogue of tunneling) from WGMs, thereby
keeping the high $Q$-factor. 
Ref.~\cite{CCBHTH94} reported unidirectional lasing from a vertical double-disk
structure. Unfortunately, the study was restricted to the near-field pattern. 
Another suggestion has been to introduce a linear defect into the
evanescent inner region of WGMs~\cite{AR04}. 
Nearly spherical, high-$Q$ fused-silica cavities showed emission into four
directions explained by {\it dynamical tunneling} from a WGM
to the exterior of the cavity~\cite{LWFN03}. 
Dynamical tunneling is a generalization of conventional tunneling which allows
to pass not only through an energy barrier but also
through other kinds of dynamical barriers in phase space~\cite{DH81}. 

In this Letter, we overcome the trade-off between $Q$-factor and
directionality by combining dynamical tunneling and refractive escape. We
couple a uniform high-$Q$ mode (HQM) and a directional low-$Q$ mode (LQM) using enhanced dynamical tunneling near {\it avoided resonance
  crossings} (ARCs).   
ARCs are generalizations of avoided frequency (or energy level) crossings. 
The latter occurs when the spectrum of a closed system is changed under 
variation of an external parameter. ARCs appear in open systems, where a
complex frequency $\omega-i\Delta\omega/2$ is assigned to each mode.  
Two types of ARCs can be distinguished~\cite{Heiss00}. The strong coupling 
situation exhibits a frequency repulsion and a linewidth crossing upon which 
the eigenstates interchange their identity. Correspondingly, all
spatial mode characteristics such as, e.g., the far-field patterns
switch their identity. The weak coupling situation consists of a frequency
crossing and a linewidth repulsion. Here, the eigenstates, and also
the spatial mode characteristics, do not interchange but only intermix
near the crossing point. Moreover, the $Q$-factors are maintained. 
Our idea is to exploit this mechanism to ''hybridize'' a HQM and a directional 
LQM to a mode with high $Q$-factor and the directed far-field pattern of the
LQM. 
This idea can be realized in three steps. First, take a cavity with
HQMs, e.g., a microdisk.
Second, introduce a one-parameter family of perturbations such 
that at least one HQM is almost unaffected and at least one HQM turns into a 
LQM having directed  emission via refractive escape. Third, 
vary the parameter such that an ARC occurs between the HQM and the LQM.
This scheme allows the systematic design of modes with high $Q$-factors and
highly directed emission. 

\begin{figure}[t]
\includegraphics[width=0.6\figurewidth]{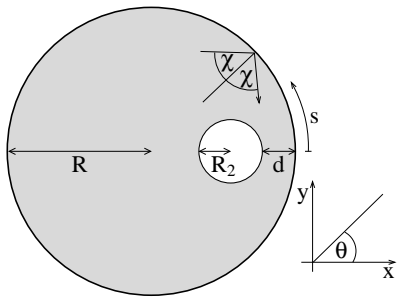}
\caption{Schematic top view of the annular cavity with radius $R$. The radius
  of the hole is $R_2$, the distance to the disk's circumference is $d$. The
  direction of the outgoing light is characterized by the angle $\theta$.}
\label{fig:annulus}
\end{figure}

We demonstrate the applicability of this scheme by a theoretical
study of a semiconductor microdisk with a circular air hole as
illustrated in \fig~\ref{fig:annulus}.  
Different versions of such an annular cavity have been studied
in the context of quantum chaos~\cite{HR02}, optomechanics~\cite{SWH04} and 
dynamical tunneling~\cite{HN97}.  
The closed system with perfectly reflecting walls has
been used as model for dynamical tunneling~\cite{BBEM93,FD98}.   

Holes with radii $\geq100$nm can be pierced through the disk surface by 
techniques currently applied to photonic crystal
membranes~\cite{BHADHPI05}. In Ref.~\cite{BCHBK99} square-shaped holes had
been introduced into a GaAs disk to reduce the laser threshold by
perturbing the nonlasing modes. 
Our calculations apply to a small GaAs microdisk of radius $R=1\mu$m. 
Similar disks have been used recently to demonstrate strong coupling
between excitons and photons in single quantum dots~\cite{PSMLHGB05}. We
choose a slab thickness of $375$nm, a temperature of 4K and a free-space
wavelength $\lambda$  close to $900$nm. This results in an effective index of
refraction $\neff = 3.3$ for the transverse magnetic polarization with 
electric field perpendicular to the disk plane. We focus on this polarization,
the conclusions are the same for transverse 
electric polarization.   

We use the boundary element method~\cite{Wiersig02b} and the $S$-matrix
approach~\cite{HR02} to compute the spatial profile of the electromagnetic
field, the $Q$-factors and the normalized frequencies $\Omega = \omega R/c$,
$c$ being the speed of light in vacuum.    
According to the discrete symmetry, the modes are divided
into even and odd symmetry classes. 

\begin{figure}[t]
\includegraphics[width=0.875\figurewidth]{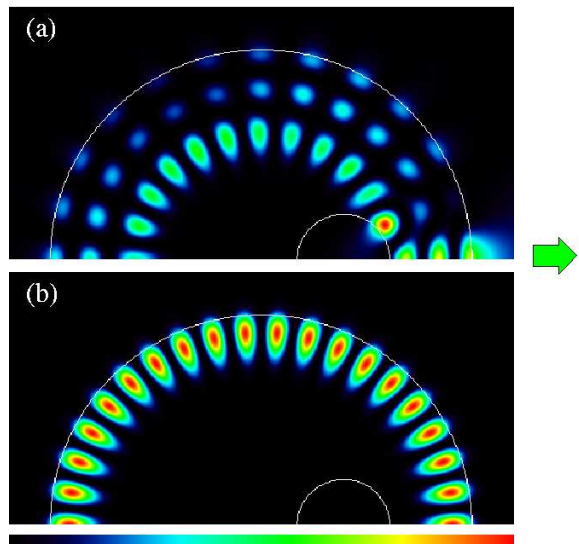}
\caption{(Color online) (a) Calculated near-field intensity pattern 
for LQM, $\Omega = 7.0599$, $Q \approx 300$, even symmetry, $R_2/R =
0.22$ and $d/R = 0.389$. The arrow shows the direction of light emission. 
(b) WGM$_{19,1}$, $\Omega = 7.0278$, $Q \approx 553000$.
}
\label{fig:modes}
\end{figure}

A WGM in a disk without hole is characterized by azimuthal and radial
mode numbers $(m,n)$. The introduction of a hole modifies a
WGM$_{m,n}$ the stronger the larger $n$ is. This can be seen from the mode in
\fig~\ref{fig:modes}(a) which bears a faint resemblance to a WGM$_{12,3}$
but is a LQM with directed emission due to refractive escape.
In contrast, a WGM$_{m,n}$ with $n = 1$ is nearly unaffected, see
\fig~\ref{fig:modes}(b). Such modes have high $Q$-factors and are expected to
emit without any preferred direction~\cite{AR04}.    
Here, we encounter the trade-off between $Q$-factors and directionality in a
single cavity.

\begin{figure}[t]
\includegraphics[width=0.95\figurewidth]{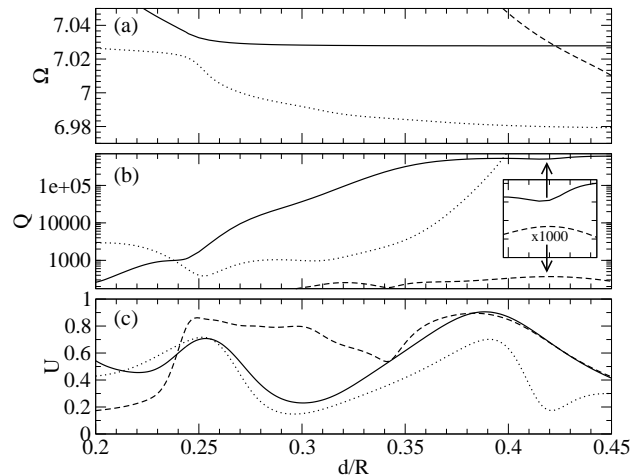}
\caption{Normalized frequencies (a) and $Q$-factors (b) of even-symmetry modes
 vs. $d$ at fixed $R_2/R = 0.22$. Inset, $Q$s in linear
 scale near the linewidth repulsion in the interval
 $[0.4,0.44]$. The low-$Q$ branch (dashed) is multiplied by 1000. 
(c) Fraction $U$ of emitted light into $|\theta| \leq 40^\circ$ for the same
 modes as in (a) and (b).}  
\label{fig:avoidedcrossing}
\end{figure}

We turn now to the realization of an ARC as we vary the parameter $d$, the 
minimal distance of the hole to the disk's boundary. 
Figure~\ref{fig:avoidedcrossing}(a) shows the
normalized frequencies of all even-symmetry modes with $\Omega\in [6.97,7.05]$ corresponding to $\lambda \approx 900$nm. This 
figure reveals the typical scenario. For $d/R > 0.35$, WGMs such as
WGM$_{19,1}$ (solid line) and WGM$_{15,2}$ (dotted) are much less affected
by moving the hole than LQMs (dashed).  
With decreasing $d/R$ all frequencies blueshift since dielectric 
material is effectively removed from the WGM region. This results in 
ARCs between WGMs and LQMs, see \figs~\ref{fig:avoidedcrossing}(a) and (b).
At $d/R\approx 0.24$ a frequency repulsion together with a linewidth 
($Q$-factor) crossing takes place. At
$d/R\approx 0.42$ we observe a frequency crossing and a linewidth repulsion (cf. inset).

\begin{figure}[t]
\includegraphics[width=0.75\figurewidth]{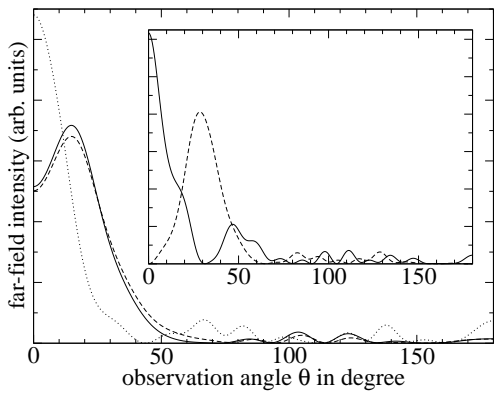}
\caption{Angular dependence of the far-field intensity for
  WGM$_{19,1}$ (solid line) and LQM (dashed) from \fig~\ref{fig:modes}. Dotted
  line marks WGM$_{34,1}$, $d/R = 0.399$, $\Omega = 11.8833$.
  Inset, even (solid) and odd (dashed) symmetry WGM$_{36,1}$, $d/R =
  0.4478$, $\Omega = 12.5241$.
}  
\label{fig:farfield}
\end{figure}

Figure~\ref{fig:avoidedcrossing}(c) shows the directionality $U$ versus $d$.
We define $U$ as the fraction of light emitted into an angular range of 
$\pm 40^\circ$ around $\theta = 0^\circ$. 
Remarkably, near the frequency crossing at $d/R \approx 0.423$ the
directionality of the WGM$_{19,1}$ and of the LQM behave in exactly the same
way.  
Gradually decreasing $d/R$, the directionality of the WGM remains close to
that of the LQM until it reaches the maximum at $d/R \approx 0.389$, where
$90\%$ of the light is emitted into a range of $\pm
40^\circ$. Figure~\ref{fig:farfield} shows that at this value of $d$ the 
angular emission pattern of the WGM and the LQM are almost identical.   
The far-field pattern has a narrow beam divergence angle of about $57^\circ$
(full width at half maximum). The spectacular finding is that a WGM with
ultra-high $Q \approx 553000$ possesses highly directed emission.  
The explanation is the hybridization of modes
near the ARC. Note that even a small contribution of the LQM to the
WGM dominates the far-field pattern because of the strong leakiness of 
the LQM.

Two additional features in \fig~\ref{fig:avoidedcrossing}(c) fit naturally 
into this scheme. First, the directionality of WGM$_{15,2}$, a spectator of
the ARC at $d/R \approx 0.42$, is less affected by the LQM. Second, near the 
ARC of the WGMs at $d/R \approx 0.24$ the directionalities of both (strongly
modified) WGMs are similar.  

\begin{figure}[t]
\includegraphics[width=0.825\figurewidth]{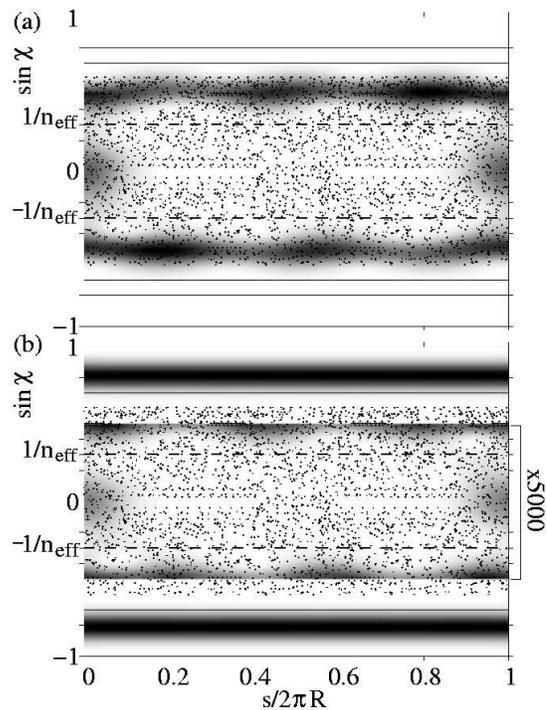}
\caption{Poincar\'e section (dots, solid lines)
  and Husimi function (shaded regions)  of modes in
  \fig~\ref{fig:modes}. 
(a) LQM. (b) WGM$_{19,1}$, the intensity inside $|\sin{\chi}|
\leq 1/2$ is multiplied by 5000 to allow for a comparison with the 
LQM.
The dashed lines mark the leaky region.  
}
\label{fig:Husimi}
\end{figure}

To see the relation to dynamical tunneling we have to compare mode patterns 
with ray dynamics. A common approach in the field of quantum chaos is to 
superimpose the Husimi function (smoothed Wigner function) of a mode onto the
Poincar\'e section~\cite{HSS03} as shown in \fig~\ref{fig:Husimi}.
The Poincar\'e section restricts the phase space of rays to the outer boundary
of the cavity. The remaining variables are $(s,\sin{\chi})$ where $s$ is the 
arclength along the circumference of the disk and $\chi$ is the angle of
incidence measured from the surface normal; cf. \fig~\ref{fig:annulus}. A good
visualisation of the dynamics is achieved by reflecting the rays a finite
number of times in the closed cavity with perfectly
reflecting outer boundary~\cite{HR02}. 
The phase space of the annular cavity contains two regular regions defined by 
$|\sin{\chi}| > 1-d/R$. A ray 
with such a large angle of incidence never encounters the hole. It rotates
as whispering-gallery trajectory ($\sin{\chi} =$ const) clockwise, $\sin{\chi}
> 0$, or counterclockwise, $\sin{\chi} < 0$. 
In particular, it cannot enter the region 
$|\sin{\chi}| < 1-d/R$. For small $R_2/R$, this region is dominated by
chaotic trajectories hitting the hole several times and thereby filling
this phase space area in a pseudo-random fashion. Embedded into the 
chaotic region is the leaky region $|\sin{\chi}|<1/\neff$. In the open cavity,
a ray entering this region escapes refractively according to Snell's and
Fresnel's laws.

The Husimi function of the LQM is restricted to the chaotic part of phase
space, see \fig~\ref{fig:Husimi}(a). The significant overlap with the leaky
region explains the low $Q$-factor.  
In contrast, the WGM lives mainly in the regular region, see
\fig~\ref{fig:Husimi}(b),  
far away from the leaky region ensuring a high $Q$. 
However, a small contribution exists in the chaotic region.
To highlight this contribution, the intensity inside the
region $|\sin{\chi}| \leq 1/2$ is multiplied by 5000. The
striking similarity to the LQM in \fig~\ref{fig:Husimi}(a) carries over 
to their far-field patterns in \fig~\ref{fig:farfield}.  

The role of tunneling is studied through the time evolution of a pure WGM 
which we define as a wave packet having no contribution in the chaotic
region. Such a mode is a superposition of the real (hybridized) WGM and the 
LQM in \fig~\ref{fig:Husimi} such that the LQM cancels the chaotic
contribution of the real WGM.   
As time evolves, the initial cancellation vanishes 
since the LQM component in the superposition is short lived. In other words, a 
fraction of the initial pure WGM has tunnelled from the regular to 
the chaotic region which allows for directed emission.     

\begin{figure}[t]
\includegraphics[width=0.9\figurewidth]{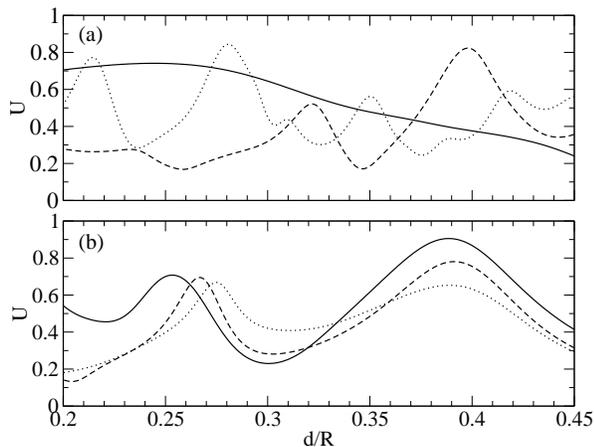}
\caption{(a) Directionality vs. $d$ for various frequencies $\Omega$ at fixed
$R_2/R = 0.22$.
WGM$_{15,1}$ with $\Omega \approx 5.71$ (solid line), 
WGM$_{34,1}$ with $\Omega \approx 11.88$ (dashed), 
WGM$_{38,1}$ with $\Omega \approx 13.16$ (dotted).
(b) WGM$_{19,1}$ for various hole radii: $R_2/R = 0.22$ (solid), $0.28$
(dashed), and $0.315$ (dotted).}   
\label{fig:directionality}
\end{figure}

Figure~\ref{fig:directionality}(a) shows the directionality for various
normalized frequencies $\Omega$. It illustrates that for lower (higher) 
frequency less (more) peaks appear. This is fully consistent with the
fact that the larger the frequency the higher is the density of
modes and therefore the larger the number of ARCs. 

The best emission directionality we found for WGM$_{34,1}$ with 
$\Omega \approx 11.88$, which corresponds to $R =
1.7\mu$m at $\lambda \approx 900$nm or $R = 1\mu$m at $\lambda
\approx 530$nm. The maximum in $U$ is at $d/R = 0.399$, see
\fig~\ref{fig:directionality}(a). The  
far-field pattern is shown in \fig~\ref{fig:farfield} as dotted line. The
emission is unidirectional with an angular divergence of $28^\circ$ 
which is much smaller than the values reported for rounded
triangles ($90^\circ$)~\cite{KLRK04} and spiral-shaped disks
($60^\circ$)~\cite{KTMJCC04}, and  
only $50\%$ larger than that of a VCSEL-micropillar of the same
radius~\cite{RBGG01}. The $Q$-factor is so huge that our
numerics is pushed to the limit; as a conservative lower bound we estimate $Q
\approx 10^8$. In practise, this theoretical bound for the $Q$-factor is
unreachable nowadays due to absorption~\cite{MLSGPL92} and surfaces 
roughness~\cite{RZ03}, i.e. the introduction of the hole does not affect the
experimental $Q$-factor.   

Figure~\ref{fig:directionality}(b) reveals that the directionality depends only
 weakly on the hole radius. This behaviour allows rather large
 fabrication tolerances without compromising the directionality much.

Let us now briefly turn to the odd symmetry modes. Since the ARC scenario is
different for even and odd symmetry, the respective far-field patterns will
differ. For low $n$ WGMs, the introduction of a hole only slightly lifts the 
even-odd degeneracy present in a perfect disk.
Hence, a light emitter couples easily to both WGMs which possibly
leads to less directed emission.  
This problem is resolved in the following cases: 
(i) In realistic microdisks, the degeneracy is lifted
due to Rayleigh scattering from the boundary~\cite{BSBP04}. 
(ii) For a single-photon source it is desirable to place a single 
emitter (e.g. a quantum dot) on an antinode of an optical mode in order to
enhance light-matter interaction. In such a case, the emitter does not couple
to the other-symmetry mode which has zero intensity there.  
(iii) It is possible to find pairs of ultra-high-$Q$ WGMs showing
simultaneously highly directed output, see inset in \fig~\ref{fig:farfield}. 
In this particular case, the far-field maxima are displaced
from each other which is interesting for applications like all-optical
switches and sensors.  

Finally, the angular divergence of the output beam in vertical direction can be
estimated to be $2/\sqrt{m}$ as in the case of an ideal disk~\cite{LL97}. For
our examples with $m=19$, 34, 38 follow angular divergences of $26.3^\circ$,
$19.7^\circ$, $18.6^\circ$.

In summary, we presented a new and general concept to achieve unidirectional
light emission from high-$Q$ modes in microcavities utilizing
resonance-enhanced 
dynamical tunneling. For a pierced microdisk, we showed  
unidirectional emission from modes with ultra-high $Q > 10^8$. 

We would like to thank T.~Harayama, S.~Shinohara, S.~Sunada, H.~Lohmeyer,
N.~Baer, and F.~Jahnke for discussions. J.W. acknowledges financial support by
the Deutsche Forschungsgemeinschaft. 

\bibliographystyle{prsty}
\bibliography{}

\end{document}